\documentclass[11pt,twoside]{atmp}

\usepackage{amsmath,amssymb}

\usepackage[all]{xy}
\usepackage{color}
\numberwithin{equation}{section}

\copyrightnotice{2007}{1}{25}{48} 

\begin{document}

\title[A New Formulation of General
Relativity] {A New Formulation of General Relativity - Part II:
Pre-Radar Charts and Generating Functions in Arbitrary
Space-Times}

\arxurl{hepreference}

\author[ Joachim Schr\"oter]{ Joachim Schr\"oter}

\address{Department of Physics, University of Paderborn,\\ D-33095 Paderborn, Germany}  
\addressemail{J@schroe.de}

\begin{abstract}
{\small{In this paper (Part II), the so-called inverse problem is
treated. This means the question whether it is possible to define
pre-radar charts, i.e.\ generating functions in arbitrary
space-times. This problem is subtle. A local general constructive
solution of it is presented. Sufficient conditions for the
existence of global solutions are given.}}
\end{abstract}

\maketitle

\section{Generating Functions}

\subsection{General Characteristics}

{\bf 7.1.1:} In Part I a function $\Psi$ generating charts was
introduced within the theory $\Phi_R$ (or $\Phi ^\ast_R$) by
Definition (3.7), and in Proposition (5.15) it was stated that
$\Psi$ also generates $g$ and  $v$. In the present Part II the
question is to be treated whether the concept of a generating
function, i.e. a family of pre-radar charts, is definable only
with the help of some special axioms of the theory $\Phi_R$ (or
$\Phi^\ast_R$) or whether it can be introduced also for arbitrary
space-times. The conditions for a positive answer
will be given in Chapter 8. In Chapter 7 we do some necessary preliminary work.\\

\cutpage

{\bf 7.1.2:} The goal of this section is summarizing {\it all} the
properties $\Psi$ has and to define the concept of a generating
function also within relativistic theories which are  different
from $\Phi_R$ and $\Phi^\ast_R$. In order to do so, at first the
conditions are written down which later on will be seen to be satisfied by $\Psi$.\\

Let $(M, \mathcal{A}^+)$ be a connected Hausdorff  $C^k$-manifold,
$k \ge 3$. Moreover, let $g$ be a Lorentzian metric and $v$ be a
velocity field on $(M, \mathcal{A}^+)$, i.e.  $g (v, v) = -1$, and
let $g$ and $v$ be of class $C^r, r \ge 2$. Then the Conditions P1
to P5
governing a term $\Psi$ read as follows:\\

\textit{\bf P1:} $\Psi$ is a $C^k$-function, $k \ge 3$,

\begin{equation} \label{P1} 
\Psi: \bigcup \limits_{q \in M} V_q \times \{q\} \rightarrow
\mathbb{R}^4
\end{equation}

with $V_q$ an open subset of $M$ and $q \in V_q$ such that
${\mathcal{A}}= \{(V_q, \Psi (\cdot, q)) : q \in M \}$ is a
$C^k$-atlas,
$k \ge 3$ which is $C^k$-compatible with ${\mathcal{A}}^+$.\\

\textit{\bf P2:} $\Psi$ generates $g$, i.e.\\

\begin{equation} \label{P2} 
g (p) = \eta_{\kappa \lambda} d_p  \Psi^\kappa (p, q) |_{q=p}
\otimes \left. d_p \Psi^\lambda (p, q) \right|_{q=p}
\end{equation}
for all $p \in M$. \\

\textit{\bf P3:} $\Psi$ generates $v$ , i.e.

\begin{equation} \label{P3} 
v (p)  = \left.\partial_{\Psi^4 (p, q)} \right|_{q=p}.
\end{equation}

for all $p \in M$.\\

\textit{\bf{P4:}} For each $q \in M$   let $\gamma_q: J_q
\rightarrow M, J_q \subset \mathbb{R}$ be a solution of
$\dot\gamma_q = v(\gamma_q)$  such that there is a $t_q \in J_q$
for which $\gamma_q (t_q) = q$. Then $J_q$ is an interval and

\begin{equation} \label{P4}  
\gamma_q (t) = \Psi (\cdot, q)^{-1} (0, 0, 0, t)
\end{equation}

for all $t \in J_{q}$. Moreover, $ J_{q}$ is also the domain of the right-hand side of (7.4).\\

\textit{\textbf{ P5:}} For all $q' \in W_{q} : =  \text{ran} \gamma_{q}$ the equation $\Psi (\cdot, q) = \Psi ( \cdot, q')$ holds.\\

In other words, the function $\Psi$ generates an atlas
$\mathcal{A}$ of pre-radar charts, a metric $g$, a velocity field
$v$ and the integral curves
$\gamma_q$ of $v$. Loosly speaking, $\Psi$ knows almost everthing one is interested in GR.\\

Then the following proposition is true. \\
{\bf Proposition (7.1):} The function $\Psi$ introduced within the
theory  $\Phi_R$ (or $\Phi^\ast_R$) by Definition (3.7) satisfies
the
Conditions P1 to P5.\\

\textit{\bf Proof}: Since $\Psi (p, q) = \hat{\Psi} (F (q), p) =
\psi_A (p)$ for each $q \in W_A$ the Conditions P1, P4 and P5
follow
directly from the Axioms GK 1, 2 and 5. The Conditions P2 and P3 are satisfied because of Proposition (5.15).\\

{\bf 7.1.3:} These considerations suggest generalizing the concept
of
a generating function also to relativistic theories different from $\Phi_R$ and $\Phi^\ast_R$ as follows.\\

{\bf Definition (7.2)}: 1. Let $(M, {\mathcal{A}}^+)$ be a
$C^k$-manifold, $k \ge 3$, for which $g$ is a Lorentzian metric
and $v$ a velocity field. Then any term $\Psi$ satisfying the
Conditions P1 to P5 is called a (full) generating function. If
$\Psi$ satisfies P1 and perhaps some, but not all of the
conditions P2 to P5 it is called a partial generating function. If
$\Psi$ satisfies only P1 and P2 the coordinates $\Psi (\cdot, q)$
are known as locally Minkowskian.\\
2. If $\Psi$ generates a $C^k$-atlas ${\mathcal{A}}, k \ge 3$, and
if $\mathcal{D}$ is the differential structure of class $C^k$
containing $\mathcal{A}$
we say that $\mathcal{D}$ is generated by $\Psi$. \\
3. The coordinates generated by a full generating function are called pre-radar coordinates. \\

{\bf Remark (7.3):} If $\Psi$ is a partial generating function
which satifies P1 and P 4 then it satisfies also P3. For from P4
we have the equations

\begin{equation} \label{rem7.3}  
\Psi (\gamma_q (t), q) = (0, 0, 0, t)
\end{equation}

and $\dot\gamma_q (t) = v (\gamma_q (t))$ for each $t  \in  J_q$.
Furthermore, it follows from (\ref{rem7.3}) that

\begin{equation} \label{rem7.3-1} 
\dot\gamma_q (t) = \left.\frac{d}{dt} \Psi^\alpha (\gamma_q (t),
q) \; \partial_{{\Psi^{\alpha}} (\cdot, q)} \right|_{\gamma_q (t)}
= \left.\partial_{\Psi^4 (\cdot, q)}  \right|_{\gamma_q (t)}.
\end{equation}

Therefore, for each $q  \in M$ we find

\begin{equation} \label{rem7.3-2} 
\left.v(q) = v(\gamma_q (t_q)) = \dot\gamma_q (t_q) =
\partial_{\Psi^4 (\cdot, q)}  \right|_{q}
\end{equation}

so that (\ref{P3}) holds.\\
 Hence P3 in Definition (7.2) is superfluous, but there are practical reasons to take P3 as a separate condition.
This will be seen e.g.\\ in Section 7.2. \\

{\bf Remark (7.4):} 1. From Condition P5 one concludes that
$\gamma_q = \gamma_{q'}$ for each $q' \in W_q: = $ ran $\gamma_q$.
Similarly we have $J_q = J_{q'}$ and $W_q = W_{q'}$ for $q' = W_q$. 
\begin{sloppypar}
2. It follows from Condition P4 that for each $q \in M$ the
function $\Psi (\cdot, q)^{-1} (0, 0, 0, \cdot)$ is an integral
curve of $v$.\\  
\end{sloppypar}
3. Condition P1 has the consequence that any two partial
generating functions have a commun domain because $q \in V_q$, and
that they generate the same differential structure $\mathcal{D}$.

\subsection{Relations between generating functions}

In this section the problem is to be treated to which extent the
Conditions P1 to P5 determine the partial generating functions. In
a first step two partial generating functions $\Psi$ and $\Psi'$
are considered which satisfy the Conditions P1 and P2. Then the
following proposition holds.\\

{\bf Proposition (7.5):} Let $\Psi, \Psi'$ satisfy P1. Moreover,
let $g$ be generated by $\Psi$. Then $\Psi'$ generates $g$ exactly
if

\begin{equation} \label{prop7.5} 
\Psi' (p, q) =  L (q) \cdot \Psi (p, q)  + R (p, q)
\end{equation}

where $L (q) = ((L^\alpha_\beta)(q)))$ is a Lorentz matrix and
$d_p R (p, q) \mid_{q=p} \; = 0$. (Here $\Psi'$, $\Psi$ and $R$
are column vectors.) \\

{\bf Proof:} Let $\Theta^{'\kappa} = d \Psi^{'\kappa} \mid_{q=p}$
and $\Theta^\alpha = d \Psi^\alpha \mid_{q = p}$. Then by
definition we have $g = \eta_{\alpha \beta} \Theta^\alpha \otimes
\Theta^\beta$. Moreover, let $g' : = \eta_{\kappa \lambda}
\Theta^{'\kappa} \otimes \Theta^{'\lambda}$. Now assume
(\ref{prop7.5}) to be valid. Then $\Theta^{'\kappa} =
L_\alpha^\kappa \Theta^\alpha$ and because of
$\eta_{\alpha \beta} = \eta_{\kappa \lambda} L^\kappa_\alpha L^\lambda_\beta $ we have $g = g'$. \\
Conversely, let $g$ and $g'$ be as above and assume $g' = g$.
Moreover, let $e_{\beta}$ denote the duals of
$\Theta^{\alpha}$.Then $g ( e_\alpha, e_\beta) = \eta_{\alpha
\beta} = \eta_{\kappa \lambda} \Theta^{'\kappa} (e_\alpha)
\Theta^{'\lambda} (e_\beta)$. Therefore, the matrix $L =
((L_\alpha^\kappa))$ with $L_\alpha^\kappa = \Theta^{'\kappa}
(e_\alpha)$ is a Lorentz matrix. Since $L^\kappa_\alpha =
L^\kappa_\varrho \Theta^\varrho (e_\alpha)$ one concludes that

\begin{equation} \label{proof7.5} 
\Theta^{'\kappa} = L^\kappa_\varrho \Theta^\varrho.
\end{equation}

This equation reads explicitly
$\left. d \Psi^{'\kappa} (p, q) \right|_{q=p} = \left. L^\kappa_\varrho (p) d \Psi^\varrho (p, q) \right|_{q = p}.$\\

Hence $d (\Psi^{'\kappa} (p, q) - L_\varrho^\kappa (q)  \Psi^\varrho (p, q)) \mid_{q=p} = 0 $ so that (\ref{prop7.5}) results.\\

{\bf Remark (7.6):} If (\ref{prop7.5}) is true the elements
$L^\kappa_\alpha$ of the Lorentz matrix $L$ are given by
$L^\kappa_\alpha =
 \Theta^{'\kappa} (e_\alpha)$. Hence this equation holds if $\Psi'$ and $\Psi$ both generate the metric $g$. Therefore,
in each case, $L^\kappa_\alpha$ is of class $C^r, r \ge 2$.
Consequently, $R (p, \cdot)$ is of class $C^r, r \ge 2$,  too.
Mixed derivatives and derivatives with respect to  $p$ alone exist up to third order.\\

In the next step we consider two partial generating functions
which satisfy the Conditions P1, P2 and P3. Clearly the result of
Proposition (7.5) holds. But it turns out that the Lorentzian matrix $L$ has a more special form.\\

{\bf Proposition (7.7):} Let $\Psi, \Psi'$ satisfy P1. Moreover
let $\Psi$ generate $g$ and $v$, and let $v'$ be a velocity field.
Then $\Psi'$   generates $g$ and $v'$ exactly if $\Psi'$ is given
by (\ref{prop7.5}) where $L$ satisfies the condition

 \begin{equation} \label{prop7.7} 
v'^{\flat} = - L^4_\alpha \Theta^\alpha.
\end{equation}

{\bf Proof:} If $\Psi'$ generates $g$ and $v'$ the Relation
(\ref{prop7.5}) is valid and $v'^{\flat} = -\Theta^{'4}$.
Moreover, Equation (\ref{proof7.5}) holds because $\Psi'$
generates $g$. Because of $v'^{\flat} = - \Theta^{'4}$ Equation
(\ref{prop7.7}) holds, too. Now suppose the converse to be true.
If (\ref{prop7.5}) holds then $\Psi'$ generates $g$. Therefore
there is a Lorentz matrix $L$ for which
Equation (\ref{proof7.5}) is valid. If Equation (\ref{prop7.7}) is true we find that $v'^{\flat} = -\Theta^{'4}$. Hence $\Psi'$ generates $g$ and $v'$.\\

{\bf Corollary (7.8):} The unique solution of (\ref{prop7.7}) is

\begin{equation} \label{corol7.8} 
L^4_\alpha = - v'^{\flat} (e_\alpha).
\end{equation}

Therefore (\ref{prop7.7}) is equivalent to (\ref{corol7.8}). Since
$v'^{\flat} (e_\alpha) = v'_\alpha$ at any point $q \in M$ are the
covariant components of $v'$ in $\Psi (\cdot, q)$ - coordinates we
have $\eta^{\alpha \beta} v'_\alpha v'_\beta = -1$.

This equation has the consequence that there is a Lorentz matrix
$L$ such that (7.10) holds. The proof of this statement can be
read off
from the parts 3 and 5 of the proof of Proposition (8.2), where a somewhat more general case is treated.\\

{\bf Remark (7.9):} Since the velocitiy fields $v$ and $v'$ define
time orientations, they define the same orientation if $g (v, v')
< 0 $.
In this case $L^4_4> 0$ so that finally $L^4_4 \ge 1$. For $g (v', v) = v'^{\flat} (v) = - L^4_4$ because $v = e_4$.\\

{\bf Corollary (7.10):} Let the suppositions be as in Proposition
(7.7). Then $\Psi'$ generates $g$ and $v$ exactly if $\Psi'$ is
given by (\ref{prop7.5}) with

\begin{equation} \label{corol7.10} 
L = \left(
\begin{array}{cc}
Q  & 0  \\
0 & 1
\end{array} \right)
\end{equation}

where $Q(q)$ is an orthogonal $3 \times 3$ matrix for each $q \in M$.\\

{\bf Proof:} Let (\ref{prop7.5}) and (\ref{corol7.10}) be valid.
Then $g=g'$ and

\begin{equation} \label{proof7.10} 
\Psi^{'4} = \Psi^4 + R^4
\end{equation}

Therefore, $v'^\flat = - \Theta'^4 = -\Theta^4 = v^\flat$ so that
$v' = v$. Now let $g = g'$ and $v = v'$. Since $g'$ and $v'$ are
generated by $\Psi'$,  Equation (\ref{corol7.8}) holds. It reads
in this case: $L^4_\alpha = \Theta^4 (e_\alpha) = \delta^4_
\alpha$.
Then it follows from general properties of Lorentz matrices that $L$ has the form (\ref{corol7.10}).\\

Now let $\Psi$ and $\Psi'$ be partial generating functions
satisfying P1, P2 and P3. What can be said about $\Psi'$ if
in addition $\Psi$ satisfies P4?  The answer is given by \\

{\bf Proposition (7.11):} Assume that $\Psi$ and $\Psi'$ satisfy
Condition P1. Moreover, assume that $\Psi$ generates $g$ and $v$
and that  it satisfies P4 where $\gamma_q$ is related to $\Psi$ by
(\ref{P4}). Then $\Psi'$ generates $g$ and $v$, and satisfies
Condition P4 exactly if $\Psi'$ is given by (\ref{prop7.5}) and
(\ref{corol7.10}) where the additional condition

\begin{equation} \label{prop7.11} 
R (\gamma_q (t), q) = 0
\end{equation}

holds for all $t \in J_q$.\\

{\bf Proof:} First assume that $\Psi'$ generates $g$ and  $v$, and
satisfies P4. Hence, in P4 the integral curves with respect to
$\Psi'$ are the same as with respect to $\Psi$. Then, because
$\Psi'$ and $\Psi$ are related by (\ref{prop7.5}) together with
(\ref{corol7.10}), we find

\begin{equation*}
(0, 0, 0, t)^T = \Psi' (\gamma_q (t), q) = {\left
(\begin{array}{cc}
Q(q)  & 0  \\
0 & 1
\end{array} \right)}
\cdot (0, 0, 0, t)^T + R (\gamma_q (t), q),
\end{equation*}

so that Equation (\ref{prop7.11}) holds.

Conversely, if $\Psi'$ is given by (\ref{prop7.5}) and
(\ref{corol7.10}) it generates $g$ and $v$ so that the integral
curves in
 P4 with respect to $\Psi'$ and $\Psi$ are the same. Therefore

\begin{equation*}
\Psi' (\gamma_q (t), q) = {\left(
\begin{array}{cc}
Q(q) & 0\\
0 & 1
\end{array}\right)}
\cdot (0, 0, 0, t)^T + R (\gamma_q (t), q)
\end{equation*}

so that condition P4 for $\Psi'$ is satisfied because of (\ref{prop7.11}).\\

A similar result holds if one takes condition P5 into account. At first some suppositions are specified.\\

It is supposed that $\Psi$ and $\Psi'$ satisfy Condition P1.
Moreover, it is assumed that $\Psi$ generates $g$ and $v$, and
that for each $q \in M$ there is an integral curve $\gamma_q$ of
$v$ where $J_q : =$ dom $\gamma_q$ is an interval and where there
is a $t_q  \in J_q$ with $\gamma_q (t_q) = q.$
Finally it is supposed that $\Psi$ satisfies P5 with $W_q: =$ ran $\gamma_q$.\\

{\bf Proposition (7.12):} If these assumptions are true then
$\Psi'$ generates $g$ and $v$, and satisfies P5 exactly if $\Psi'$
is given by (7.8) and (7.12) where the additional conditions

\begin{equation} \label{prop7.12} 
Q (q') = Q (q) \quad \text{and} \quad R (\cdot, q') = R (\cdot, q)
\end{equation}

hold for all $q' \in W_q$.\\

{\bf Proof:} If $\Psi'$ is given by (\ref{prop7.5}) and
(\ref{corol7.10}) then it generates $g$ and $v$. In addition, if
(\ref{prop7.12}) are true $\Psi'$ satisfies also P5. Conversely,
if $\Psi'$ generates $g$ and $v$, it is given by (\ref{prop7.5})
and (\ref{corol7.10}). If in addition $\Psi'$ satisfies P5 the
equations

\begin{equation*}
\frac{d}{dt} \Psi (p, \gamma_q (t)) = 0, \quad   \frac{d}{dt}
\Psi' (p, \gamma_q (t)) = 0
\end{equation*}

hold for all $p \in V_q$. Hence one concludes from (\ref{prop7.5})
that

\begin{equation*}
\Psi^\beta (p, \gamma_q (t)) \frac{d}{dt} L^\alpha_\beta (\gamma_q
(t)) + \frac{d}{dt}  R^\alpha (p, \gamma_q (t)) = 0. \nonumber
\end{equation*}

Since $R$ is a least of class $C^2$ one can apply the d-Operator
with respect to $p$ at the point $p = \gamma_q (t)$. Then we
obtain

\begin{equation} \label{proof7.12} 
\Theta^\beta (\gamma_q (t)) \frac{d}{dt} L^\alpha_\beta  (\gamma_q
(t)) = 0.
\end{equation}

For $\alpha = 4$ it is an identity, and for $\alpha = l = 1,2,3$
it reads

\begin{equation} \label{proof7.12-1} 
\Theta^j (\gamma_q (t)) \frac{d}{dt} Q^l_j  (\gamma_q (t)) = 0.
\end{equation}

Since $\Theta^j, j=1, 2, 3$ are linearly independent we find
$Q(q')= Q(q)$ for all $q' \in W_q$. Inserting this result into
(\ref{prop7.5}) and (\ref{corol7.10}) we obtain also the second
part of (\ref{prop7.12}).

\section{Construction of Generating Functions}

\subsection{The inverse problem}

In the Chapters 3 and 4 (of Part I) an axiomatic formulation of
the frame theory $\Phi_R$ is given (and analogously for
$\Phi^\ast_R$ in Chapter 6) which is based on the existence of
pre-radar charts. The set of all these charts forms a generating
function for the metric $g$ and the velocity $v$. Since in the
usual formulation of GR the existence of pre-radar charts or of a
generating function is not postulated, the question arises whether
it is possible to construct a generating function in this case. I
call
this question the {\it inverse problem}.\\

In order to formulate the problem precisely let us first describe
a relativistic (frame) theory $\Phi^+$ which represents the usual
account of GR. The base sets of $\Phi^+$ are the set of (signs
for) events $M$ and the reals $ \mathbb{R}$ (and possibly other
sets). The structural terms are an atlas ${\mathcal{A}}^+$, a
metric $g$, a velocity $v$, a mass density $\eta$ and an empirical
temperature $\vartheta$ (and possibly other fields). In any case
the axioms of $\Phi^+$ contain geometrical and kinematical axioms,
the equations of motion of matter, Einstein's equation and
additional conditions. In vacuum theories the density of matter is
zero so that the equations of motion are empty and the right-hand
side of Einstein's equation is zero.
Then the invers problem can be stated thus:\\

{\bf Problem (8.1):} 1. If $\Phi^+$ is given, is there a (full)
generating function $\Psi$ in $\Phi^+$ in the sense of Definition
(7.2)?\\
2. If $\Phi^+$ is given, is there a local solution to the problem
in the following sense: there is a covering of $M$ by open sets
$V$ so that
a generating function $\Psi$ exists for each open submanifold with base set $V$?\\

This problem can be solved in two ways: first one shows that a
solution exists, and second, one constructs an explicit term which
represents a generating function for each given theory $\Phi^+$.
In what follows I shall present a general constructive solution of
the local problem. But in addition, this result allows to
formulate sufficient conditions for solutions of the global
problem.

\subsection{Construction of a field of tetrads}

{\bf 8.2.1:} In this section we assume that a theory of type
$\Phi^+$ is given. More specific, we consider a $C^k$-manifold
$(M, {\mathcal{A}}^+), k\ge 3$ with a Lorentz metric $g$ and a
velocity $v$ defined on $M$ such that $g (v, v) = -1$ and such
that $g$ and $v$ are of class $C^r, r \ge 2$. The construction of
a local (full) generating function in Section 8.3 is based on the
existence of a tetrad field the components of which determine the
components of $g$ and $v$ in the sense of Formulae (5.1) and
(5.2). For this purpose let us take a chart $(V, \chi) \in
{\mathcal{D}}$ where ${\mathcal{D}}$ is the differential structure
of class $C^k, k \ge 3$ which contains  ${\mathcal{A}}^+$,
and let $g_{\alpha \beta}$ and $v^\kappa$ be the $\chi$-components of $g$ and $v$. Then the following result holds.\\

{\bf Proposition (8.2):} For each $x \in \chi [V]$ there is a
matrix $\Lambda (x) = ((\Lambda_\sigma^\varrho (x)))$ with det
$\Lambda (x) \not= 0$ and such that

\begin{equation} \label{prop8.2}
g_{\alpha \beta} (x) = \Lambda^\kappa_\alpha (x)
\Lambda^\lambda_\beta  (x) \eta_{\kappa \lambda}
\end{equation}

and

\begin{equation} \label{prop8.2-1}
v^\kappa (x) = \Lambda^{-1}\left.^\kappa_4\right. (x).
\end{equation}

Moreover, the covectors $\Theta^\alpha (p) = \Lambda^\alpha_\beta
(x) d x^\beta, \alpha = 1. \cdots, 4$ and the vectors $e_\kappa
(p) = \Lambda^{-1}\left.^\lambda_\kappa\right. (x)
\partial_{x^\lambda}, \kappa = 1, \cdots, 4$ with $x = \chi (p), p
\in V$ form orthogonal tetrads in $T^\ast_p V$ resp. in $T_p V$.

The {\bf proof} is effected in  five steps.\\
1. Since the matrix $((g_{\alpha p}))$ is nonsingular, symmetric
and of signature 2 there is an orthogonal matrix
$((Q^\varrho_\sigma))$ such that $g_{\alpha \beta} =
Q^\varrho_\alpha Q^\sigma_\beta d_{\varrho \sigma}$ where

\begin{equation*}
d_{\varrho \sigma} = {\text{diag}} (a_1^2, a_2 ^ 2, a_3^2, -
a_4^2)_{\varrho \sigma}, \quad a_\lambda > 0, \quad  \lambda=1,
\cdots, 4 .
\end{equation*}

Now let $f^\nu_\mu = \sum\limits^4_j \delta^\nu_j  \delta^j_\mu
a_j$. Then $d_{\varrho \sigma} = f^\kappa_\varrho f^\lambda_\sigma
\eta_{\kappa \lambda}$. Defining $K: = ((K^\kappa_\alpha))$  by
$K^\kappa_\alpha = Q^\varrho_\alpha f^\kappa_\varrho$ we obtain
the relations det $K \not= 0$ and

\begin{equation} \label{prop8.2-2}
g_{\alpha \beta} = K^\kappa_\alpha K^\lambda_\beta \eta_{\kappa
\lambda}.
\end{equation}

2. From (8.1) and (8.3) it follows that $K$ and the matrix
$\Lambda$ we are looking for, can differ at most by a Lorentz
matrix $L$, i.e.

\begin{equation} \label{prop8.2-3}
\Lambda ^\kappa_\alpha = L^\kappa_\varrho  K^\varrho_\alpha  \; .
\end{equation}

Since (8.2) is equivalent to

\begin{equation}   \label{8.2}
\Lambda^4_\alpha = - g_{\alpha \kappa} v^\kappa
\end{equation}

the next task is to find a Lorentz matrix $L$ such that

\begin{equation} \label{8.2-1}
L^4_\varrho K^\varrho_\alpha = - g_{\alpha \kappa} v^\kappa
\end{equation}

for the given terms $K^\varrho_\alpha, g_{\alpha \kappa}$ and
$v^\kappa$. The unique solution of (8.6) reads simply

\begin{equation} \label{8.2-2}
L^4_\beta = -K^{-1}\left.^\alpha_\beta\right.  g_{\alpha \kappa}
v^\kappa = : h_\beta  \;  .
\end{equation}

But up to now it is not clear whether the four components
$h_\beta$ are possible candidates for the fourth row of a
Lorentz matrix. It is shown in the next three steps that such Lorentz matrix exists.\\
3. In this step a special Lorentz matrix is introduced. By simple
calculations one can see that the following equations hold:

\begin{equation} \label{8.2-3}
h_\beta = -\eta_{\beta \sigma} K^\sigma_\kappa v^\kappa,
\end{equation}

\begin{equation} \label{8.2-4}
\eta^{\alpha  \beta} h_\alpha h_\beta = -1.
\end{equation}

>From (8.9) we get $| h_4 | \ge 1$. Since both K and -K satisfy
Equation (8.3), because of (8.8) we can choose $h_4 \ge 1$. Now
let $r : = - (h^2_4 -1)^{\frac{1}{2}}$ and $\bar{v} :=(1-
h^{-2}_4)^{\frac{1}{2}}$. Because of $h_4 \ge 1$ we get $\bar{v}
\in [0, 1[$. This means that $\bar{v}$ is a speed parameter.
Moreover, by a simple calculation we find that

\begin{equation*}
h_4 = (1- \bar{v}^2)^{-\frac{1}{2}}, \,  r = - \bar{v}  (1-
\bar{v}^2)^{-\frac{1}{2}} \;   .
\end{equation*}

Thus, the matrix

\begin{equation} \label{8.2-5}
S = \left( \begin{array}{cccc}
h_4 &0 & 0 & r \\
0 & 1& 0 & 0\\
0& 0& 1& 0\\
r &0& 0&  h_4
\end{array} \right)
\end{equation}

is a special Lorentz matrix.\\
4. In this step two orthogonal matrices are defined. Let

\begin{equation} \label{8.2-6}
b^1_l : = - (h^2_4 - 1)^{-\frac{1}{2}} h_l, \quad  l = 1, 2, 3,
\end{equation}

then $\sum\limits^{3}_{l} (b^1_l)^2 = 1$. Moreover, let $B^1 =
(b^1_1, b^1_2, b^1_3)$ and in addition let $B^\varrho =
(b^\varrho_1, b^\varrho_2, b^\varrho_3)$,  $\varrho = 2, 3$ be
vectors such that $\{B^1, B^2, B^3 \}$ is an orthonormal basis in
$\mathbb{R}^3$. Then define the orthogonal matrix $B$ by $B =
(B^1, B^2, B^3)^T$. Finally, let A be an arbritary orthogonal $3
\times 3$ matrix and define $\hat{A}, \hat{B}$ by

\begin{equation} \label{8.2-7}
\hat{A} = \left(
\begin{array}{cc}
A & 0\\
0 & 1
\end{array} \right), \quad
\hat{B} = \left(
\begin{array}{cc}
B & 0\\
0 & 1
\end{array} \right) \; .
\end{equation}

5. In the last step of the proof let us consider the matrix $L =
\hat{A} \cdot S \cdot \hat{B}$. By definition, $L$ is a Lorentz
matrix and

\begin{equation} \label{8.2-8}
L^4_\beta = S^4_1  \hat{B}^1_\beta + S^4_4 \hat{B}^4_\beta \; .
\end{equation}

Inserting $S^4_1 = r$, ${\hat{B}}^1_\beta = r^{-1}
\sum\limits^{3}_{l} h_l  \delta^l_\beta$, $S^4_4 = h_4$ and
$\hat{B}^4_\beta = \delta^4_\beta$ we obtain $L^4_\beta =
h_\beta$. Hence a matrix $\Lambda$ satisfying (8.1) and (8.2) is
given by $\Lambda = L \cdot K$. Then it is easily seen that the
dual tetrads
$\Theta^\alpha, \alpha = 1, \cdots, 4$ and $e_\kappa, \kappa = 1, \cdots, 4$ are orthogonal.\\

{\bf Corrolary (8.3):} It is seen from the proof that there is not
only one matrix $\Lambda$, but there are infinitely many $\Lambda$
which generate $g_{\alpha \beta}$ and $v^\kappa$ by (8.1) and
(8.2). Moreover, if there is a Lorentz matrix $L$ such that
$\Lambda = L \cdot K$ satisfies (8.1) and (8.2) for a given $K$,
then $L = \hat{A} \cdot S \cdot \hat{B}$ where $\hat{A}, S$ and
$\hat{B}$ are the matrices introduced by (8.10) and (8.12). This
can be seen quite easily starting with a general Ansatz  $L =
\tilde{A} \cdot  \tilde{S} \cdot \tilde{B}$
and showing that  $\tilde{S} = S, \tilde{B}^1_\beta = \hat{B}^1_\beta$ and $\tilde{A}$ is any matrix of the form $\hat{A}$ in (8.12).\\

{\bf Remark (8.4).} In what follows we have to differentiate
$\Lambda$. If, as usual, $g_{\alpha \beta}$ and $v^\kappa$ are of
class $C^r$, $r \ge 2$ then the eigenvalues and eigenvectors af
$g_{\alpha \beta}$ are of class $C^r, r \ge 2$ (cf. \cite{kato66},
p. 122; due to a private communication of H. Sohr, Paderborn the
result of Kato can be generalized to the present case.) Hence the
matrix $K$ is of class $C^r, r \ge 2$, and consequently $h_\beta,
\beta = 1 \ldots, 4$ and $b^1_l, l = 1, 2, 3,$ too. Then the
vectors $B^\varrho, \varrho = 2, 3$ can be adjusted to form an
orthonormal basis together with $B^1$, so that the matrix $B$ is
of class $C^r, r \ge 2$. Finally,
let $A$ be any sufficiently smooth field of orthogonal matrices. Then $L$ and $\Lambda$ are of class $C^r, r\ge  2$.\\

{\bf 8.2.2.:} In this section as well as in the subsequent ones
the chart $(V, \chi)$ is specialised. It is assumed that $\chi$ is
a comoving
coordinate system with respect to $v$. This supposition has the following consequences.\\

{\bf Remark (8.5):} By definition of $\chi$ we get for the
components $v^\alpha$ of the velocity: $v^\alpha = \delta^\alpha_4
w$. Since $g_{44} w^2 = -1$, both $g_{44}$ and $w$ are unequal
zero, and by a special choice of $\chi^4$ we can attain the
relation $w > 0$. Moreover, since
$\Lambda^{-1}\left.^\alpha_4\right. =  v^\alpha  =
\delta^\alpha_4$ $w$ we obtain the equation $\Lambda^{-1}
\left.^4_4\right. = w$ whereas the other elements of the fourth
column of $\Lambda$ are zero. Since
$\Lambda = (\Lambda^{-1})^{-1}$, a similar result holds for $\Lambda$, i.e. $\Lambda^\alpha_4 = \delta^\alpha_4 w^{-1}$.\\

In a further step the integral curves of $v$ are determined. First of all we introduce\\

{\bf Notation (8.6):} 1. Let, as above, $\tilde{V} = \chi [V] =
\mathrm{ran} \chi$. Then define
\begin{equation} \label{not8.6}
\bar{V}: = \{(x^1, x^2, x^3): \;\text{there is an} \; x^4 \;
\text{such that} \; (x^1, \cdots, x^4) \in \tilde{V}\}.
\end{equation}

If $x = (x^1, \cdots x^4)$ the abbreviations $\bar{x} = (x^1, x^2, x^3)$ and $x = (\bar{x}, x^4)$ are used. \\
2. If $z \in \tilde{V}$ and if $w$ is as in Remark (8.5) then $f$
is defined for a fixed $x^4_0$ by

\begin{equation} \label{not8.6-1}
f (\bar{z}, x^4) = \int\limits^{x^4}_{x^4_0} \frac{d \xi}{w
(\bar{z}, \xi)}.
\end{equation}

3. For each $\bar{z} \in \bar{V}$ let

\begin{eqnarray} \nonumber
T_{\bar{z}}: & = & \{x^4 : (\bar{z}, x^4) \in \tilde{V} \} \\
\nonumber J_{\bar{z}}:& = & \{ \tau:  \; \text{there is an} \; x^4
\in T_{\bar{z}} \; \text{and} \;  \tau = f (\bar{z}, x^4)  \}.
\nonumber
\end{eqnarray}

{\bf Remark (8.7):} In what follows it is assumed that there is an
$x^4_0$ such that $\bar{V} \times \{ x^4_0 \} \subset \tilde{V}$.
By restriction of $V$ such $x^4_0$ can always be found. Then $f$
is defined with exactly one of these numbers $x^4_0$. For the same
reason we may assume that $T_{\bar{z}}$ and $J_{\bar{z}}$ are
intervals. Because of $w > 0$ the function $f (\bar{z}, \cdot)$ is
strictly increasing and of class $C^{r+1}, r \ge 2$. Therefore the
inverse function exists and is also of class
$C^{r+1}, r \ge 2$. We write $\varphi (\bar{z}, t) := f (\bar{z}, \cdot)^{-1} (t)$.\\

Then the following proposition holds.\\
{\bf Lemma (8.8):} The path $\gamma$ is a solution of

\begin{equation} \label{lem8.8}
\dot\gamma^\alpha = v^\alpha (\gamma) = \delta^\alpha_4 w (\gamma)
\end{equation}

exactly if there is a $\bar{z} \in \bar{V}$ such that

\begin{equation} \label{lem8.8-1}
\gamma^l (t) = z^l, \quad l = 1, 2, 3,
\end{equation}

\begin{equation} \label{lem8.8-2}
\gamma^4 (t) =  \varphi (\bar{z}, t)
\end{equation}

for all $t \in J_{\bar{z}}$.\\

{\bf Proof:} If $\gamma$ is a solution of (\ref{lem8.8}), then
$\gamma^l, l = 1, 2, 3$ is constant, and by separation of
variables one gets (8.17). The converse holds because
$\dot\gamma^l (t) = 0, l = 1, 2, 3$ and

\begin{displaymath}
\dot\gamma^4 (t) = \frac{\partial}{\partial t} \varphi (\bar{z},
t) = w (\bar{z}, \gamma^4 (t)).
\end{displaymath}

To complete the tools needed in the next section we introduce the\\

{\bf Notation (8.9):} If $\gamma^l (t) = z^l, l = 1, 2, 3,$ we
write $\gamma = \gamma_{\bar{z}}$ and $W_{\bar{z}} = \mathrm{ran}
\gamma_{\bar{z}}$. Sometimes it is convenient to write also
$\gamma_{z'}: = \gamma_{\bar{z}}$ and $W_{z'}: = W_{\bar{z}}$ for
$z' \in W_{\bar{z}}$.

\subsection{Solution of the local inverse problem}

{\bf 8.3.1 Suppositions:} In what follows we consider again a
relativistic theory of type $\Phi^+$ which is characterized by the
base sets $M$ and $\mathbb{R}$, the structural terms
${\mathcal{A}}^+$, $g$ and $v$ and possibly others, and the axioms
ruling these terms (and possibly
others). (Cf. the Sections 8.1 and 8.2.1.)\\
Let us assume that a comoving chart $(V, \chi)$ with respect to
the velocity $v$ is given. Then the Notation (8.6) and (8.9) is
used
as well as the results of Remark (8.7) and Lemma (8.8).\\
Finally, it is assumed that the construction of the tetrad
components $\Lambda^\alpha_\beta$ determining $g$ and $v$ by (8.1)
and (8.2) is carried through in the chart $(V,  \chi)$. But with
respect to differentiability we assume a stronger condition to be
valid: the components
$\Lambda^\alpha_\beta, \alpha,  \beta = 1, \cdots, 4$ are of class $C^k, k \ge 3$.\\

{\bf 8.3.2:} At first a function $\tilde{\Phi}$ is defined. Later
on it turns out that a restriction of it determines the generating
function we are
looking for. In the definition Notation (8.6) is used.\\

{\bf Definition (8.10):} The function

\begin{equation}  \label{def8.10}
\tilde{\Phi}: \bigcup\limits_{z \in \chi [V]} (\mathbb{R}^3 \times
T_{\bar{z}}) \times \{z\} \rightarrow \mathbb{R}^4
\end{equation}

is defined by

\begin{equation} \label{def8.10-1}
\begin{array}{cc}
\tilde{\Phi}^j (x, z)  =  \sum\limits^3_l (x^l - z^l) \Lambda^j_l (\bar{z}, x^4), \;  j = 1, 2, 3,  \\
\tilde{\Phi}^4 (x, z) =  \sum\limits^3_l (x^l - z^l) \Lambda^4_l
(\bar{z}, x^4) + f (\bar{z}, x^4).
\end{array}
\end{equation}

Hence $\tilde{\Phi} \in C^k, k \geq 3$ if $\Lambda^\alpha_\beta$, $\alpha, \beta = 1,..., 4$ are of class $C^k, k \geq 3.$\\
Before proofing that a restriction of $\tilde{\Phi}$ is a generating function two lemmas have to be proved.\\

{\bf Lemma (8.11):} For each $z \in \chi [V]$ there is an open set
$V_{\bar{z}} \subset \mathbb{R}^3 \times T_{\bar{z}}$ such that
$\tilde\Phi (\cdot, z)$ restricted to $V_{\bar{z}}$ is bijective and $\{\bar{z}\} \times T_{\bar{z}} \subset V_{\bar{z}}$.\\

{\bf Proof:} 1. Let $y = \tilde{\Phi} (x, z)$ and $X = (x^1  - z^1, \cdots, x^3 - z^3, w(\bar{z}, x^4) f (\bar{z}, x^4))$.\\
Then (\ref{def8.10-1}) is equivalent to $ \Lambda^{-1} (\bar{z},
x^4) \cdot y^T  = X^T,$ i.e.

\begin{equation} \label{lem8.11}
\sum\limits^3_l \Lambda^{-1}\left.^j_l \right. (\bar{z}, x^4) y^l
+ z^j = x^j, \quad  j= 1, 2, 3
\end{equation}

\begin{equation} \label{lem8.11-1}
\Lambda^{-1}\left.^4_\alpha \right. (\bar{z}, x^4) y^\alpha = w
(\bar{z}, x^4) f (\bar{z}, x^4).
\end{equation}

Now assume that Equation (\ref{lem8.11-1}) is solvable for $x^4$.
This means there is a function $F (\bar{z}, \cdot)$ defined on a
set $U_{\bar{z}} \subset \mathrm{ran} \tilde{\Phi} (\cdot, z)$
such that $x^4 = F(\bar{z}, y)$ satisfies (\ref{lem8.11-1}) if $y
\in U_{\bar{z}}$. Then $\tilde{\Phi} (\cdot, z)^{-1}$ exists. It
is defined for all $y \in U_{\bar{z}}$ by

\begin{equation} \label{lem8.11-2}
\sum\limits^3_l \Lambda^{-1} \! \left.^i_l\right.  (\bar{z}, F
(\bar{z}, y)) y^l + z^i = x^i, i = 1, 2, 3,
\end{equation}

\begin{equation} \label{lem8.11-3}
F (\bar{z}, y) = x^4 \; .
\end{equation}

Hence, the only thing we need to prove is that Equation (\ref{lem8.11-1}) is solvable for $x^4$ if $y$ is an element of a certain set $U_{\bar{z}}$.\\
2. It is helpful to simplify the notation. In this and the next
steps of the proof the fixed vector $\bar{z}$ is omitted. Since
$\Lambda^{-1} \! \left.^4_4\right. = w > 0$, Equation
(\ref{lem8.11-1}) reads

\begin{equation} \label{lem8.11-4}
\sum\limits^3_l \sigma_l (x^4) y^l + y^4 = f (x^4)
\end{equation}

where $\sigma_l = w^{-1} \Lambda^{-1} \! \left.^4_l\right.$. Since
$f$ is bijective by Remark (8.7) one defines $\varrho_l (t)  =
\sigma_l (f^{-1} (t))$, so that Equation (\ref{lem8.11-4}) becomes

\begin{equation} \label{lem8.11-5}
\sum\limits^3_l  \varrho_l (t)  y^l + y^4 = t,
\end{equation}

and we arrive at the result, that (\ref{lem8.11-4}) has a unique
solution $x^4 \in T_{\bar{z}}$ exactly if
(\ref{lem8.11-5}) has a unique solution $t \in J_{\bar{z}}$.\\

Let $N_{\bar{z}} : = \{(0, 0, 0, \tau) : \tau \in J_{\bar{z}} \}$
then (\ref{lem8.11-5}) is solvable if $y \in N_{\bar{z}}$, namely
$t = y^4$. In the
next steps of the proof the solvability of (\ref{lem8.11-5}) is extended to a neighborhood $U_{\bar{z}}$ of $N_{\bar{z}}$.\\
3. Since each open interval is the union of a countable family of
increasing closed finite intervals we need to prove the
solvability of (\ref{lem8.11-5}) only for closed finite intervals.
Thus let $J_i \subset J_{\bar{z}}$ be such an interval, then
$\varrho_l$ and $\dot\varrho_l$ are bounded in $J_i$ because
$\varrho_l$ is of class $C^r, r \ge 2$, i.e. there is a number
$m_i$ such that

\begin{equation} \label{lem8.11-6}
\sum\limits^3_l \varrho_l^2 (t) < m_i \quad \text{and} \quad
\sum\limits^3_l \left| \dot{\varrho}_l (t) \right| < m_i
\end{equation}

for all $t \in J_i$. If $J_i = [a_i, b_i]$ and $0 < \zeta < 1$
define

\begin{equation} \label{lem8.11-7}
\delta_i (y^4) = \mathrm{min} \{| y^4 - a_i|, | y^4 - b_i| , \zeta
\}
\end{equation}

and

\begin{equation} \label{lem8.11-8}
U_i  =  \{ y : a_i < y^4 < b_i,  \sum\limits^3_l  (y^l)^2 <
\frac{1}{m_i} \delta_i (y^4)  \}.
\end{equation}

Then $\delta_i$ is continuous in $J_i$, and therefore $U_i$ is
open. Now let us consider the function $H$ defined by

\begin{equation} \label{lem8.11-9}
H (y, t) = \sum\limits^3_l \varrho_l (t) y^l + y^4
\end{equation}

for $y \in U_i$ and $t \in J_i$.\\
4. It is shown that for each $y \in U_i$ the function $H (y,
\cdot)$ maps $J_i$ into $J_i$ and is contracting. Let $t \in J_i$
and $t' = H (y, t)$ then

\begin{equation} \label{lem8.11-10}
|t' - y^4| \le \sum\limits^3_l \varrho_l (t)^2 \cdot
\sum\limits^3_r (y^r)^2 < \delta_i (y^4).
\end{equation}

>From (\ref{lem8.11-10}) one concludes that $t' \in J_i$. Moreover
let  $t,  t' \in J_i$, $t < t'$ and $H (y, t) =  \tau, \quad H (y,
t') = \tau'$. Then there is a $\tilde{t} \in J_i$ with $t \le
\tilde{t} \le t'$ such that

\begin{equation} \label{lem8.11-11}
\tau - \tau' = \sum\limits^3_l  \dot{\varrho}_l (\tilde{t}) y^l (t
- t')  .
\end{equation}

Hence for all $y \in U_i$

\begin{equation} \label{lem8.11-12}
|\tau - \tau'|  <  \delta_i (y^4) |t - t'| \le \zeta |t - t'|
\end{equation}

where $0 < \zeta < 1$, so that $H (y, \cdot)$ is contracting. Then
by Banach's fixed-point theorem (cf.\ e.g.\ \cite{Muel} p.\ 251,
\cite{Reed72} p.\ 151 ) we get that for each $y \in U_i$ there is
exactly one $t \in J_i$ so that the equation $H (y, t) = t$ holds.
Consequently, there is a function
$G_i$ such that $G_i (y) = t$ for each $y \in U_i$ and $G_i$ is uniquely determined.\\
5. If $y \in U_i \cap U_j, \; i \not= j$ then $G_i (y) = G_j (y)$
because $G_i$ and $G_j$ are the solutions to the same equation,
i.e. Equation (\ref{lem8.11-5}). Now we can define the function F
we are looking for (cf. (\ref{lem8.11-3})) as follows: Define
$U_{\bar{z}}$ by $U_{\bar{z}} = \bigcup\limits_i U_i$ and $G$ by
$G (\bar{z}, y) = G_i (y)$ for $y \in U_i$. Then let $F$ be
defined by

\begin{equation} \label{lem8.11-13}
F (\bar{z}, y) = f^{-1} (\bar{z}, G (\bar{z}, y))
\end{equation}

for all $y \in U_{\bar{z}}$. By construction $x^4 = F (\bar{z}, y)$ is a solution of (\ref{lem8.11-1}).\\
6. Therefore, by the considerations of step one we conclude that
$\tilde{\Phi} (\cdot, z)$ restricted to

\begin{equation}  \label{lem8.11-14}
V_{\bar{z}}: = \{x : \tilde{\Phi} (x, z) \in U_{\bar{z}}   \}
\end{equation}

is bijective. The set $V_{\bar{z}}$ is open because
$\tilde{\Phi}(\cdot, z)$ is
continuous and because $U_{\bar{z}}$ is open.\\

{\bf 8.3.3:} This lemma now enables us to define a function which
later on turns out to be the coordinate representation of the
generating
function we are looking for.\\

{\bf Definition (8.12):} The function $\Phi$ is defined by $\Phi
(x, z) = \tilde{\Phi} (x, z)$ for all

\begin{equation} \label{def8.12}
(x, z) \in \bigcup\limits_{z' \in \chi [V]} V_{\bar{z}'} \times \{
z' \} = :  \mathrm{dom}  \Phi.
\end{equation}

The second lemma we need concerns the differentiability of $\Phi (\cdot, z)^{-1}$.\\

{\bf Lemma (8.13)}: If $\Lambda \in C^k, k \ge 3$ (cf. Section 8.3.1) then $\Phi (\cdot, z)^{-1} \in C^k, k \ge 3$.\\

{\bf Proof:} To simplify notation the fixed vector $\bar{z}$ is
omitted. From (\ref{lem8.11-2}) and (\ref{lem8.11-3}) we see that
$\Phi (\cdot, z)^{-1} \in C^k$, if $F:= F(\bar{z}, \cdot)$ is of
class $C^k$. Let us write (\ref{lem8.11-4}) in the form

\begin{equation} \label{def8.12-1}
Z (y, x^4) := \sum\limits^3_l \sigma_l (x^4) y^l + y^4 - f(x^4)
\end{equation}

where $y \in U_{\bar{z}}$ and $x^4 \in T_{\bar{z}}$. Then $Z (y,
F(y)) = 0$ by definition of $F$. Since $\sigma_l = w^{-1}
\Lambda^{-1} \! \left.^4_l\right., w = \Lambda^{-1} \!
\left.^4_4\right.$, $w > 0$ the functions $\sigma_l, l = 1, 2, 3 $
and $f$ are of class $C^k$,
so that Z is also of class $C^k$.\\
Now using the notation of (\ref{lem8.11-5}) we obtain $\sigma_l
(x^4) = \varrho_l (f (x^4))$ and therefore

\begin{equation} \nonumber
\frac{\partial \sigma_l}{\partial x^4} = \frac{\partial
\varrho_l}{\partial t} \cdot \frac{\partial f}{\partial x^4} =
\dot{\varrho}_l \frac{1}{w} \; .
\end{equation}

Hence

\begin{equation} \label{def8.12-2}
\frac{\partial Z}{\partial x^4} = \frac{1}{w} \sum (\dot\varrho_l
y^l -1) \;  .
\end{equation}

For each $t \in J_{\bar{z}}$ and each $y \in U_{\bar{z}}$ there is
an $i \in {\mathbb{N}}$ such that $t = f (x^4) \in J_i$ and $y \in
U_i$. Therefore

\begin{equation*}
\left| \sum\limits^3_l \dot\varrho_l y^l \right| < \delta_i (y^4)
\leq \zeta < 1,
\end{equation*}

so that $|\frac{\partial Z}{\partial x^4}| > 0$. Applying the
Implicite Function Theorem (cf.\ e.g.\ \cite{flem64} p.\ 117) we
conclude that $F = F (\bar{z}, \cdot)$ is
also of class $C^k, k \ge 3$. Thus the proof is complete.\\

{\bf 8.3.4:} With the help of the Lemmas (8.11) and (8.13) we are now able to prove the main result of Section 8.3.\\

{\bf Proposition (8.14):} If the suppositions of Section (8.3.1)
are fulfilled and if $\Phi$ is given by Definition (8.12) then
$\Psi = \Phi (\chi, \chi)$ is a generating function in the submanifold $V$ with the global coordinate system $\chi$.\\

{\bf Proof:} 1.\ The function $\Psi$ is of class $C^k, k \ge 3$,
because $\Phi$ is of class $C^k, k \ge 3$, by the Definitions
(8.10) and (8.12). It follows from the Lemmas (8.11) and (8.13)
that for each $z \in \chi [V]$ the function $\Phi (\cdot, z)$ is a
diffeomorphism of class $C^k, k \ge 3$, i.e. it is a
transformation of coordinates. Hence for each $q \in V$ the
function $\Psi (\cdot, q)$ is a coordinate function, and the set
of all these functions is an atlas on $V$ which is compatible with the (global) chart $(V, \chi)$. Thus Condition P1 is satisfied.\\
2. By a simple calculation one can see that

\begin{equation*}
\frac{\partial \Phi^\alpha}{\partial x^\beta} (x, z) \mid_{z=x} =
\Lambda^\alpha_\beta (x).
\end{equation*}

Since by construction of $\Lambda$ the equations (5.1)
and (5.2) hold, and because of Formulae (5.9)
and (5.10) one can see
quite easily that $\Psi$ generates $g$ and $v$. Hence the Conditions P2 and P3 are fulfilled.\\
3. The integral curves $\hat{\gamma}_q$ of $v$ are determined by
the Formulae (\ref{lem8.8-1}) and (\ref{lem8.8-2}) and by
$\hat{\gamma}_q (t) = \chi^{-1} (\gamma_z (t))$ where $z = \chi
(q)$. Then for each $z = \chi (q), q \in V$ and $t \in
J_{\bar{z}}$ we have $\Psi (\hat{\gamma}_q (t), q) = \Phi
(\gamma_z (t), z) = (0, 0, 0, t)$.
Hence P4 is fulfilled.\\
4. Let $\hat{W}_q: = \mathrm{ran} \hat{\gamma}_q$. Then $\hat{W}_q
=  \chi^{-1} [W_z]$ where $z = \chi (q)$. Moreover, $\Psi (\cdot,
q') = \Psi (\cdot, q)$ for $q' \in \hat{W}_q$ exactly if $\Phi
(\cdot, z') = \Phi (\cdot, z)$ for $z' \in W_z$. But the latter
equation
is true because $\Phi$ depends only on $\bar{z}$. Therefore Condition P5 is satisfied, too.\\

{\bf Corollary (8.15):} For a given theory $\Phi^+$ of the type
described in Section 8.2.1 the velocity field $v$ is defined on
$M$. Hence, for each $q \in M$ there is a neighborhood $V$ such
that a comoving coordinate system $\chi$ exists on $V$. Therefore,
by Proposition (8.14) the second part of Problem (8.1), the local
form of the inverse problem, is solved.
If $V = M$ one has obtained a solution of the global Problem (8.1).\\

\subsection{Examples}

{\bf 8.4.1:} The simplest case for which the inverse problem can
be solved explicitly is the following: let $(M, \mathcal{A})$ be a
space-time manifold with metric $g$ and velocity $v$, and let $(V,
\chi)$ be a chart such that

\begin{equation*}
g_{\alpha \beta} (x) = \eta_{\alpha \beta} a^2_\alpha (x) , \quad
v_\alpha (x) = - \delta^4_\alpha a_4 (x)
\end{equation*}

where $x \in \chi [V]$. Then $a_\beta (x) \not= 0$ for all $x \in
\chi [V]$, so that

\begin{equation*}
\Lambda^\kappa_\alpha (x) = \delta^\kappa_\alpha a_\alpha (x),
\quad \Lambda^{-1} \left.^\mu_\beta \right. (x) = \delta^\mu_\beta
a^{-1}_\beta (x).
\end{equation*}

Therefore \\
 \centerline{$g_{\alpha \beta} = \Lambda^\kappa_\alpha \Lambda^\kappa_\beta  \eta_{\kappa \lambda}, \quad
v_\alpha = - \Lambda^4_\alpha$} \\
and

\begin{align}\Phi^j (x, z) &=  a_j (\bar{z}, x^4) (x^j - z^j),\quad  j =
1, 2, 3\label{8.40}\\
 \Phi^4 (x, z) & =  f (\bar{z}, x^4)\label{8.41}
\end{align}

where $f$ is defined by Notation (8.6).\\

{\bf 8.4.2:} These considerations can be applied to the
Robertson-Walker space-time $(M, \mathcal{A})$ with $k = 0$. Here
the same notation is used as in Subsection 8.4.1. In this case $M
= \mathbb{R}^3 \times T$ with $T = ] 0, \infty[$ and
${\mathcal{A}} = \{(M, id_M)\}$. For the global chart $(M, id_M)$
we have

\begin{equation*}
g_{\alpha \beta} (x) = K^2 (x^4) \sum\limits^3_i \delta^i_\alpha
\delta^i_\beta - \delta^4_\alpha \delta^4_\beta
\end{equation*}

where $K (x^4) \not= 0$ for all $x^4 \in T$, and $v_\alpha (x) =
-\delta^4_\alpha$. Therefore

\begin{equation*}
\Phi (x, z) = (K (x^4) (x^1 - z^1), \cdots, K (x^4) (x^3 - z^3),
x^4).
\end{equation*}

The covariant velocity $v =  -d x^4$ is solely owing to massive particles.\\

{\bf 8.4.3:} A similarly simple case is the outer Schwarzschild
space-time $(M, \mathcal{A})$ where $M = \{y \in \mathbb{R}^3 : \|
y \| > r_0\} \times \mathbb{R}$  and ${\mathcal{A}} = \{(M,
id_M)\}$. Using the (nonglobal) Schwarzschild coordinates $x = (r,
\vartheta, \varphi, t)$ we have

\begin{eqnarray*}
a^2_1 (x) &=& (1- \frac{r_0}{r})^{-1} , \quad  a^2_2 (x) = r^2\\
a^2_3 (x) &=& r^2 \; \text{sin}^2 \vartheta , \quad  a^2_4 (x) =
1- \frac{r_0}{r}.
\end{eqnarray*}

Since the outer Schwarzschild solution is a vacuum solution the
velocity field can be chosen such that the set of testparticles is
comoving. This means:

\begin{equation*}
v_\alpha (x) = - \delta^4_\alpha (1- \frac{r_0}{r})^{\frac{1}{2}}
.
\end{equation*}

Now let $z = (r', \vartheta', \varphi', t')$. Then the result is

\begin{eqnarray*}
\Phi^1 (x, z) &=& (1 - \frac{r_0}{r'})^{-\frac{1}{2}} (r- r'),\\
\Phi^2 (x, z) &=& r' (\vartheta - \vartheta'),\\
\Phi^3 (x, z) &=& r'  \text{sin} \vartheta' (\varphi - \varphi'),\\
\Phi^4 (x, z) &=& (1 - \frac{r_0}{r'})^{\frac{1}{2}} t.
\end{eqnarray*}

{\bf 8.4.4:} The last result can be generalized. For each vacuum
solution $g$ which is diagonal in a chart $\chi$ a generating
function $\Psi$ is given by (\ref{8.40}) and (\ref{8.41}).
Especially if $\chi$ is a global chart then $\Psi$ is a global
solution of Problem (8.1).\\

{\bf Acknowledgement}  \\

I want to thank Mr. Gerhard Lessner for valuable discussions and
critical remarks and Mr. Wolfgang Rothfritz for correcting my
English.\\

\bibliographystyle{my-h-elsevier}

\end{document}